# Self-Aligned van der Waals Heterojunction Diodes and Transistors

*Vinod K. Sangwan,[1,#] Megan E. Beck,[1,#] Alex Henning,[1] Jiajia Luo,[1] Hadallia Bergeron,[1] Junmo Kang,[1] Itamar Balla,[1] Hadass Inbar[1], Lincoln J. Lauhon,[1] and Mark C. Hersam[1,2,3,*]*

[1]Department of Materials Science and Engineering, Northwestern University, Evanston, Illinois 60208, USA.

[2]Department of Chemistry, Northwestern University, Evanston, Illinois 60208, USA.

[3]Department of Electrical Engineering and Computer Science, Northwestern University, Evanston, Illinois 60208, USA.

[#]These authors contributed equally

*E-mail: m-hersam@northwestern.edu

**ABSTRACT**

A general self-aligned fabrication scheme is reported here for a diverse class of electronic devices based on van der Waals materials and heterojunctions. In particular, self-alignment enables the fabrication of source-gated transistors in monolayer $MoS_2$ with near-ideal current saturation characteristics and channel lengths down to 135 nm. Furthermore, self-alignment of van der Waals p-n heterojunction diodes achieves complete electrostatic control of both the p-type and n-type constituent semiconductors in a dual-gated geometry, resulting in gate-tunable mean and variance of anti-ambipolar Gaussian characteristics. Through finite-element device simulations, the operating principles of source-gated transistors and dual-gated anti-ambipolar devices are elucidated, thus providing design rules for additional devices that employ self-aligned geometries.

For example, the versatility of this scheme is demonstrated via contact-doped MoS$_2$ homojunction diodes and mixed-dimensional heterojunctions based on organic semiconductors. The scalability of this approach is also shown by fabricating self-aligned short-channel transistors with sub-diffraction channel lengths in the range of 150 nm to 800 nm using photolithography on large-area MoS$_2$ films grown by chemical vapor deposition. Overall, this self-aligned fabrication method represents an important step towards the scalable integration of van der Waals heterojunction devices into more sophisticated circuits and systems.

**TOC IMAGE**

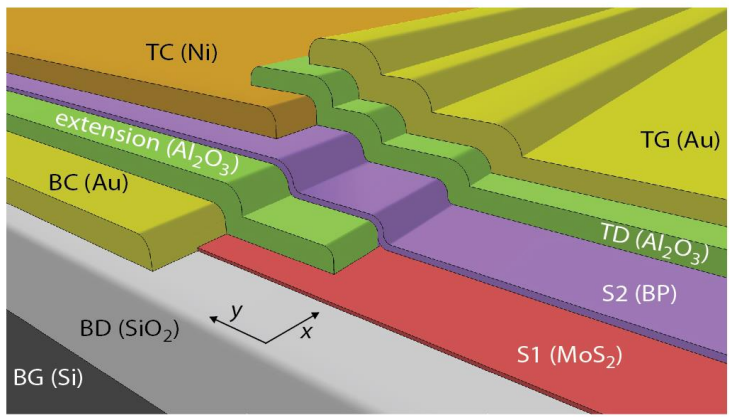

Parallel self-aligned fabrication methods in modern silicon-based microelectronics have enabled sub-lithographic registration between processing steps, ultimately facilitating substantial advances in circuit complexity over the past few decades.[1] In contrast, while two-dimensional (2D) materials have shown significant potential for digital and analog electronics due to their high mobilities, ultrathin geometry, and broad range of permutations in van der Waals heterojunctions (vdWHs),[2-9] 2D material devices have not yet exploited parallel self-aligned fabrication to achieve both short channels and large area fabrication. Thus far, short-channel 2D material transistors and vdWHs have been achieved using serial processing methods such as electron-beam lithography or mechanical placement on nanotube or nanowire gates.[5, 10, 11] Similarly, the relative alignment of different layers in vdWHs has been inhibited by the diffraction-limited resolution of transfer and alignment methods. Here, we overcome these limitations by introducing a self-aligned processing methodology that enables the fabrication of 2D material transistors with channel lengths below 150 nm with minimal short-channel effects and improved current saturation, as demonstrated with monolayer $MoS_2$. These self-aligned transistors show the highest output resistance at the lowest channel length reported for a 2D material, which is of interest for high-frequency current amplifiers and mixers. In vdWHs based on black phosphorus (BP) and $MoS_2$, this self-aligned approach allows dual-gate electrostatic control of the carrier density in both constituent semiconductors while concurrently achieving independent gate control of the short-channel series transistors. Since this self-aligned methodology is compatible with a diverse range of nanomaterials and can be implemented in parallel via large-area lithographic processes without sacrificing lateral spatial resolution, it is likely to impact a wide range of 2D and mixed-dimensional vdWH devices.

The underlying building block of our self-aligned method is a dielectric extension protruding from metal electrodes, which is formed by exploiting resist undercuts that are ubiquitous in lithographic processes. Both electron-beam lithography and photolithography resist undercuts have been optimized to obtain dielectric extensions in the range of 100 nm to 800 nm (Supporting Figs. S2.1, S2.2). For example, a bilayer of two poly(methyl methacrylate) electron-beam lithography resists, in which the lower molecular weight resist (resist 1, higher sensitivity to electron dose) is under the higher molecular weight resist (resist 2, lower sensitivity), results in an undercut down to 135 nm (Fig. 1a). The metal electrodes (metal 1, Fig. 1a) are obtained by directional evaporation (i.e., edges defined by resist 2), and the dielectric extension is obtained by conformal growth of a dielectric (i.e., edges defined by resist 1) by atomic layer deposition (ALD), followed by liftoff processes (Fig. 1a). When used for field-effect transistors (FETs), the dielectric extension defines the semiconducting (semiconductor 1) channel length ($L$) formed by subsequent metal evaporation (metal 2, Fig. 1a).

Using this methodology, self-aligned short-channel (SASC) MoS$_2$ FETs were fabricated on local gates (Au) on undoped Si wafers with 300 nm thick thermal oxide (Figs. 1b,c). The undoped wafers minimize parasitic capacitance, and the thermal oxide aids optical visualization of MoS$_2$ monolayers grown by chemical vapor deposition.[4,12] The gate dielectric and the dielectric extension are both 30 nm thick ALD-grown Al$_2$O$_3$. Atomic force microscopy (AFM) reveals a dielectric extension length of ~135 nm (Figs. 1d,e), and optical microscopy shows that the top electrode overlaps the bottom electrode and thus the channel (Fig. 1c). The inherent asymmetry in the resulting transistor geometry allows device operation in two biasing conditions: the *source-gated* configuration where the bottom electrode is biased and the overlapping electrode is grounded, and the *drain-gated* configuration where the overlapping electrode is biased and the bottom electrode is grounded. Source-gated biasing exhibits significantly improved current saturation (quantified by output impedance, $r_o = \partial V_D/\partial I_D$ at fixed $V_G$) at large $V_D$ compared to drain-gated biasing where $I_D$ continues to increase with $V_D$ for all $V_G$ (Figs. 1f,g). The saturation regime for the source-gated case at large $V_G$ = 6 V is nearly ideal ($g_D = 1/r_o = \partial I_D/\partial V_D$ < 10 pS) with a high $I_{on}/I_{off}$ ratio (~$10^5$) (Fig. 1h). In contrast, the drain-gated case shows channel-width-normalized $g_D$ = 0.5 µS/µm that cannot be turned off even at large $V_G$ = –7 V, resulting in poor on/off ratios (~$10^3$) (Fig. 1h).

This behavior is similar to conventional source-gated transistors (SGTs) that evolved from the staggered Schottky Barrier Transistor.[13] SGTs possess increased $r_o$ and intrinsic gain and decreased saturation drain voltage ($V_{SAT} = V_G–V_{TH}$) in comparison to standard FETs.[14, 15] The SGT geometry has so far been limited to silicon-on-insulator devices involving amorphous and polycrystalline silicon thin-film transistors (TFTs).[16] The SGT geometry alleviates short-channel effects and significantly improves current saturation behavior for high-speed amplifiers and impedance matching in radio frequency circuits. These advantages are achieved by 'field-relief' that results from the source contact overlapping the channel. Unlike conventional FETs where the depletion region is formed only near the drain contact,[17] the depletion region in SGTs forms first near the source contact at low $V_D$ biases, and another depletion region emerges near the drain contact at higher biases, resulting in nearly ideal current saturation and immunity against short-channel effects such as channel length modulation.[14, 15, 17] The device characteristics of conventional SGTs have been explained by three models: gate-induced source barrier lowering,[18] series resistance of the depletion region between source and channel,[19] and a thermionic emission-diffusion model with current injection concentrated at the edge of the source electrode.[15] However, most conventional SGTs use an amorphous or polycrystalline silicon semiconducting layer with thicknesses (~100 nm) comparable to the gate dielectric, in contrast to the 0.7 nm thick monolayer

MoS$_2$ used here. Thus, one can expect that the electrostatics and resulting charge transport of our SASC MoS$_2$ FETs are significantly different from those of previously reported SGTs.[15, 18]

To explore the operating principles of our SASC MoS$_2$ FETs, a device simulator (Sentaurus, Synopsys) was used to model carrier densities, potential distributions, and resulting charge transport for different short-channel device geometries and bias configurations without incorporating quantum effects specific to 2D materials (Fig. 2, Supporting Section S1.4).[7, 20-22] Simulated output characteristics assuming Ohmic contacts show a ~2-fold reduced $V_{SAT}$ and a ~7-fold reduced $g_D$ of 0.052 µS/µm at $V_D$ = 5 V for the source-gated configuration in comparison to the output curve of a back-gated FET ($g_D$ = 0.37 µS/µm) with symmetric electrode arrangement (Fig. 2a,b). Conversely, the output curve of the drain-gated FET does not saturate, showing ~13-fold larger $g_D$ of 5.1 µS/µm compared to the back-gated FET. The bias configuration determines whether the overlapping electrode reduces or increases the carrier density in the channel via a field-effect, as is reflected in the transfer characteristics (inset of Fig. 2b) and respective $V_{TH}$, which is larger for source-gated device operation (4.5 V) and lower for drain-gated device operation (3.0 V) relative to $V_{TH}$ = 3.7 V for the back-gated FET. Although the assumption of Ohmic contacts is sufficient to reveal how the channel potential profiles lead to different current saturation characteristics,[23] we find that thermionic emission and tunneling models of transport at the metal/semiconductor interfaces are needed to reproduce the low-bias ($V_D$ < 2 V) nonlinearity in the output characteristics for the drain-gated configuration (Fig. 2c).

Simulated energy band profiles (Fig. 2d, Supporting Figs. S2.4, S2.5) and electric field maps reveal the formation of an additional depletion region for the source-gated device, which leads to pinch-off near the source electrode. Consequently, the carrier density (*n*) in the semiconducting channel near the source contact is lower for the source-gated configuration than for the drain-gated configuration and the back-gated FET (Fig. 2e). The depletion region near the source electrode emerges in the source-gated FET for $V_D \geq 2$ V (Fig. 2f), whereas it is not observed for the back-gated FET (Fig. 2g) or under drain-gated device operation (Supporting Figs. S2.5, S2.6a). Consequently, the dielectric extension acts as the field-relief used in conventional SGTs by screening the drain field.[19] The evolution of the carrier density distribution in the $MoS_2$ channel as a function of the bias and the formation of depletion regions is shown for all devices in Supporting Fig. S2.5. Despite a smaller transconductance ($g_m = 0.5$ μS/μm versus 38 μS/μm) and a shorter $L$ (150 nm versus 250 nm), the experimental and simulated values of $r_o$ (>20 MΩ) for the SASC $MoS_2$ FETs are significantly higher than the previously reported values of $r_o \sim 32$ kΩ for short-channel back-gated $MoS_2$ FETs ($L = 500$ nm) that showed two orders of magnitude higher current density due to higher mobility (~50 $cm^2$/Vs).[24, 25] The ultimate scaling of these $MoS_2$ transistors to $L = 50$ nm further increased the current density (>50 μA/μm), but no current saturation was observed due to severe short-channel effects.[11]

The self-alignment approach also facilitates the reliable fabrication of p-n vdWHs with small footprints and unique electrostatic gating control. With previously reported fabrication methods, p-n vdWHs, whether lateral or vertical, consisted of a p-n heterojunction connected by two lateral p-type and n-type extensions (acting as FETs in series) or Schottky diodes with graphene, with the overall stack being coupled to one or two gates with alignment errors increasing with each component.[7-9, 26-31] In the lateral geometry, p-n vdWHs offer electrostatically controlled doping in the constituent semiconductors but suffer from large parasitic resistance from the lateral extensions beyond the junction region.[7-9, 28-30] On the other hand, vertical p-n vdWHs that employ a graphene electrode can achieve larger current density at the cost of defect-induced leakage currents, extraneous Schottky barriers, and electrode screening issues.[27, 30, 31] For example, fully vertical BP-MoS$_2$ and WSe$_2$-MoS$_2$ p-n vdWHs using graphene contacts show poor electrostatic control of $I_D$-$V_{TG}$ characteristics (Supporting Fig. S2.7). In contrast, our semi-vertical architecture addresses these shortcomings by minimizing $L$ in the lateral semiconductor extension and exposing the heterojunction to the applied electric field in a dual-gate geometry. Furthermore, our self-aligned method minimizes $L$ by controlling the size of the dielectric extension (~135 nm) rather than the resolution of optical alignment and transfer methods. Employing this approach, a dual-gated BP-MoS$_2$ p-n vdWH was fabricated on a Si substrate (global bottom gate (BG)) and 300 nm thermal oxide (bottom gate dielectric (BD)) using CVD-grown MoS$_2$ and mechanically exfoliated few-layer BP contacted with overlapping Au and Ni electrodes, respectively, and separated by a 35 nm thick Al$_2$O$_3$ extension layer (Figs. 3a-c, Supporting Fig. S2.2). The top gate dielectric (TD) of 30 nm ALD-grown Al$_2$O$_3$ and top gate (TG) of 50 nm Au are patterned in the same fabrication step, thus requiring no additional alignment.

Given the band alignment between BP and MoS$_2$, the dual-gated BP-MoS$_2$ vdWH shows rectifying *I-V* characteristics with a rectification ratio up to 50 (limited by the small band gap of BP ~ 0.4 eV) that can be controlled by both the top and bottom gates (Figs. 3d-f, Supporting Fig. S2.8).[17, 29, 32] The device behavior switches from a normal p-n heterojunction diode at $V_{TG}$ = 0 V to a Zener-like diode at $V_{TG}$ = 4 V with reversed rectification at room temperature (inset Fig. 3e), similar to a previously reported dual-gated WSe$_2$-MoS$_2$ p-n heterojunction diode operated at 77 K and an ion-gel gated BP-MoS$_2$ p-n heterojunction diode.[29, 33] Band-to-band-tunneling is barely visible in the upward trend in $I_D$ at a reverse bias of $V_D$ = –1 V at $V_{TG}$ = 0 V due to the small band gap of BP and thermal broadening of the Fermi-Dirac distribution at room temperature (Fig. 3e).[34] Unlike previous p-n vdWHs, this device shows anti-ambipolar transfer characteristics that can be tuned continuously by the bottom gate (Fig. 3g) as uniquely enabled by the self-aligned, semi-vertical architecture.[7-9]

Finite-element simulations elucidate how this unique vdWH geometry improves current rectification and enables tunable anti-ambipolar behavior (Fig. 4). Three architectures allow for gating of the lateral semiconducting extensions in dual-gated lateral vdWHs: architecture-(i) possesses individual gate control over each lateral extension (Supporting Fig. S2.9); architecture-(ii) utilizes both gates to control both lateral extensions (Figs. 4a-d, Supporting Figs. S10a,b and S11); architecture-(iii) allows both gates to control one of the lateral extensions while the other extension is controlled by a single gate (Fig. 4e, Supporting Fig. S2.10c). In architecture-(i), both semiconductors in the heterojunction can be driven to maximum electrostatic doping without compromising series resistances. However, these devices do not show anti-ambipolar behavior because dual-gate control of at least one of the lateral extensions is required (Supporting Fig. S2.9). Architecture-(ii) creates a trade-off between electrostatic control and series resistance (Figs. 4c,d), embodying almost all published examples of lateral vdWHs with one exception,[29] where the alignment of the two gates was limited by lithographic resolution. Architecture-(iii) offers controlled electrostatic doping without compromising gate-tunability of the heterojunction, resulting in control over all characteristics of the anti-ambipolar response.

Self-aligned BP-MoS$_2$ p-n vdWHs readily enable architecture-(iii). In particular, the BP flake is controlled only by the top gate due to screening of the bottom gate by MoS$_2$. Similarly, the region of the MoS$_2$ flake directly underneath BP is controlled only by the bottom gate (i1), but the rest of the MoS$_2$ flake (i2) is influenced by both gates (Figs. 4e,f). Control over the relative fractions of the total current along paths i1 and i2 results in gate-tunability of the anti-ambipolar response. Simulations of the BP-MoS$_2$ vdWH device using a topologically equivalent two-dimensional model are shown in Fig. 4f and Supporting Fig. S2.10c. An MoS$_2$ screening layer is embedded in the bottom gate dielectric to simulate the screening of the BP layer from the bottom gate field. This model reproduces the tunability of the anti-ambipolar transfer characteristics by restoring control of electrostatic depletion in the sub-threshold regime for BP ($V_{TG} > 3$ V in Fig. 4g). Simulated tunable rectification in the charge transport characteristics also agrees well with experimental data (Figs. 3f,4h, and Supporting Fig. S2.12).

In summary, we have demonstrated a self-aligned approach that enables scalable fabrication of short-channel FETs and vdWHs based on 2D semiconductors. The resulting geometry provides unique electrostatic control over charge transport including exceptional saturation characteristics in short-channel FETs. The current density in the saturation regime could be improved further by using higher mobility 2D materials such as black phosphorus or InSe.[35, 36] Thus, this approach opens new avenues of exploration for radio frequency amplifiers and mixers in ultimately scaled 2D devices.[2] Source-gating has been attempted in TFT applications such as active matrix displays, although the cutoff frequency in these cases suffered from longer channels and thicker semiconductor layers.[16] In contrast, the present self-alignment scheme is not only compatible with printed and/or flexible electronics where sub-micron resolution between layers could improve pixel density, but also presents benefits for high-mobility 2D materials due to channel length scaling. It should also be noted that the parasitic capacitance between gate and source/drain electrodes for high-frequency applications circuits could be minimized by integrating the present method with related fabrication techniques that have previously been used to align the edges of electrodes.[37, 38]

The self-aligned scheme presented in this manuscript also enables nearly complete tunability over the anti-ambipolar response in p-n vdWHs with potential implications for signal-processing applications such as frequency-shift keying and phase-shift keying (Supporting Fig. S2.13).[8] The gate-tunable mean and variance over the Gaussian anti-ambipolar response in self-aligned p-n vdWHs also possess utility for highly efficient image recognition algorithms in artificial neural networks.[39] This self-aligned fabrication approach can also be generalized to other device architectures such as contact-doped homojunction diodes (Supporting Fig. S2.14)[40] and mixed-dimensional vdWH heterojunctions (Supporting Fig. S2.15).[26] Furthermore, this self-aligned method is straightforwardly extended to large areas without compromising lateral spatial resolution as demonstrated by photolithographically defined SASC transistors on continuous $MoS_2$ films with sub-wavelength channel lengths (~150 nm) (Supporting Fig. S2.16). Overall, this work demonstrates a highly flexible and generalizable fabrication method with broad implications for electrostatically modulated 2D material and vdWH devices.

## ASSOCIATED CONTENT

**Supporting Information**:

Additional details on experimental methods, electrical characterization, and simulation data accompanies this paper and is available free of charge via the Internet at http://pubs.acs.org.

## AUTHOR INFORMATION

**Corresponding Authors:**

*E-mail: m-hersam@northwestern.edu

**ORCID:**

Vinod K. Sangwan: 0000-0002-5623-5285

Mark C. Hersam: 0000-0003-4120-1426

Lincoln J. Lauhon: 0000-0001-6046-3304

Itamar Balla: 0000-0002-9358-5743

**AUTHOR CONTRIBUTIONS:**

V.K.S., M.E.B., and M.C.H. conceived the idea and designed all the experiments. V.K.S. and M.E.B. optimized the fabrication process, measured, and analyzed all the device data. A.H., V.K.S., and L.J.L. conducted and analyzed Sentaurus TCAD device simulations. J.L., J.K., and H.I. assisted in device fabrication. H.B. and I.B. conducted growth of $MoS_2$. All authors wrote the manuscript and discussed the results at all stages. [#]These authors contributed equally.

**NOTES:**

**Competing financial interests**: The authors declare no competing financial interests.

**ACKNOWLEDGMENTS:**

This research was supported by the 2-DARE program (NSF EFRI-1433510) and the Materials Research Science and Engineering Center (MRSEC) of Northwestern University (NSF DMR-1720139). CVD growth of $MoS_2$ was supported by the National Institute of Standards and Technology (NIST CHiMaD 70NANB14H012). Charge transport instrumentation was funded by an ONR DURIP grant (ONR N00014-16-1-3179). H.B acknowledges support from the NSERC Postgraduate Scholarship-Doctoral Program. M.E.B. and H.B. acknowledge support from the National Science Foundation Graduate Research Fellowship. A.H. acknowledges the support of a Research Fellowship from the Deutsche Forschungsgemeinschaft (Grant no. HE 7999/1-1). This work made use of the Northwestern University NUANCE Center and the Northwestern University Micro/Nano Fabrication Facility (NUFAB), which are partially supported by the Soft and Hybrid

**FIGURES:**

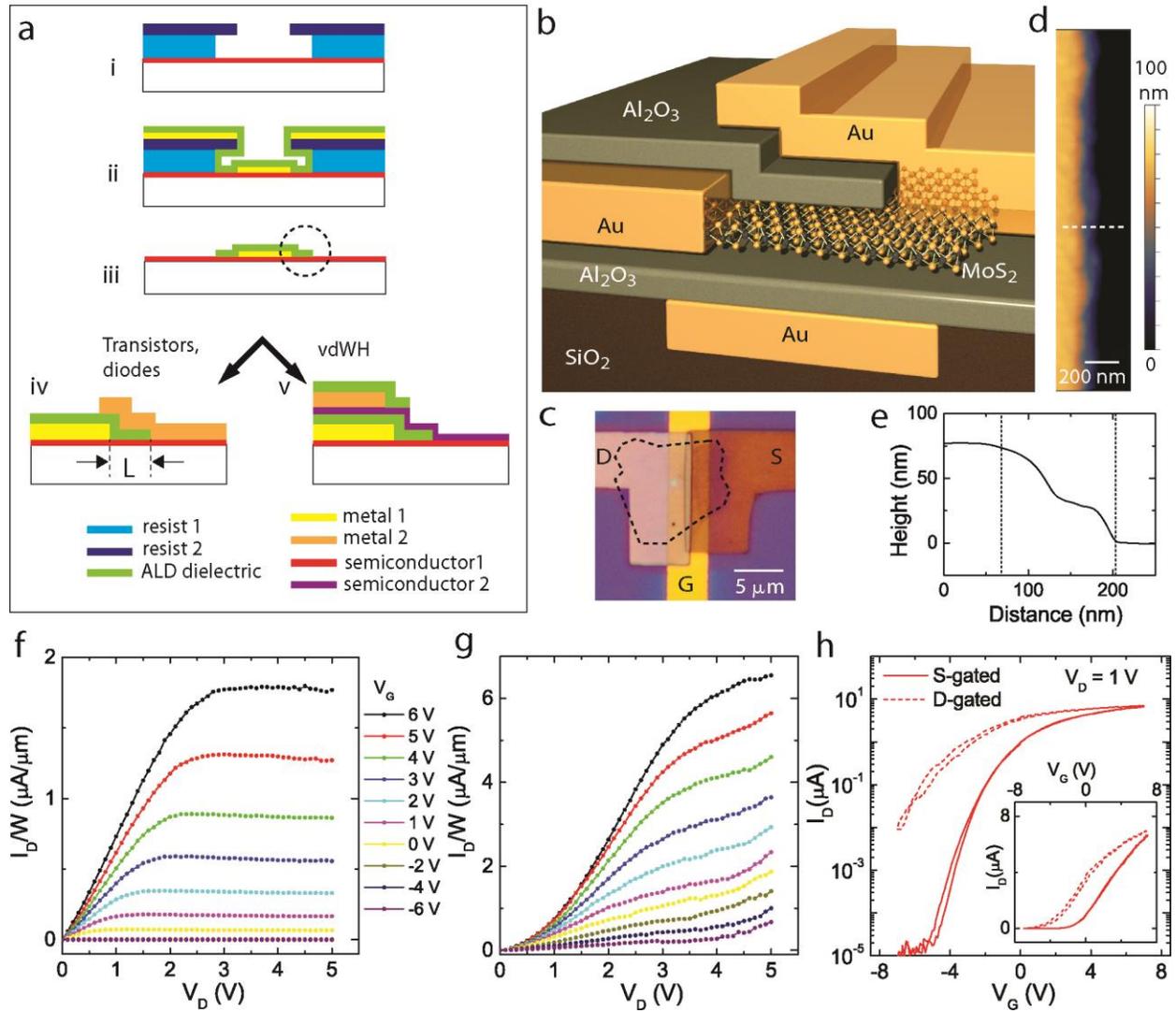

**Figure 1.** Process-flow and self-aligned short-channel MoS$_2$ transistors. (a) Fabrication scheme for a dielectric extension on a 2D semiconducting layer. (i) An undercut profile is achieved using a bilayer resist. (ii) An encapsulated metal contact is fabricated by directional evaporation of metal followed by conformal atomic layer deposition (ALD) of the dielectric followed by (iii) lift-off to remove the resist bilayer. (iv, left) A transistor or a contact-doped diode is achieved by subsequent evaporation of the same or a different metal, respectively, with the channel length (*L*) defined by the dielectric extension. (iv, right) Van der Waals heterojunctions (vdWH) are achieved by transferring another 2D semiconductor followed by metallization. See Supporting Methods S1.1, S1.2 for details. (b) Schematic of a self-aligned short-channel (SASC) MoS$_2$ field-effect transistor (FET) with a local gate on an undoped Si substrate with thermal oxide coating. The top electrode on the right-hand side overlaps the dielectric extension and thus also overlaps the channel. (c) Optical micrograph of a SASC MoS$_2$ transistor. The source-gated (S-gated) case refers to a biasing condition where the electrode under the dielectric extension (left) (i.e., the drain (D) electrode) is biased and the overlapping electrode (right) (i.e., source (S) electrode) is grounded. The drain-gated (D-gated) case employs the opposite biasing scheme. (d) Atomic force microscopy topography image of the dielectric extension (dashed circle in (a)). (e) Height profile along the white dashed line in (d) showing the dielectric extension length of ~135 nm by taking into account both vertical and horizontal growth of the ALD oxide on the metal edge. (f) Output characteristics of an S-gated SASC MoS$_2$ transistor with $L = 200$ nm showing current saturation. (g) Output characteristics of the same transistor in the D-gated configuration showing loss of current saturation. The gate bias ($V_G$) legend between (f) and (g) corresponds to both of the plots. (h) Transfer characteristics of the S-gated and D-gated cases at $V_D = 1$ V. The inset shows the same data using a linear scale.

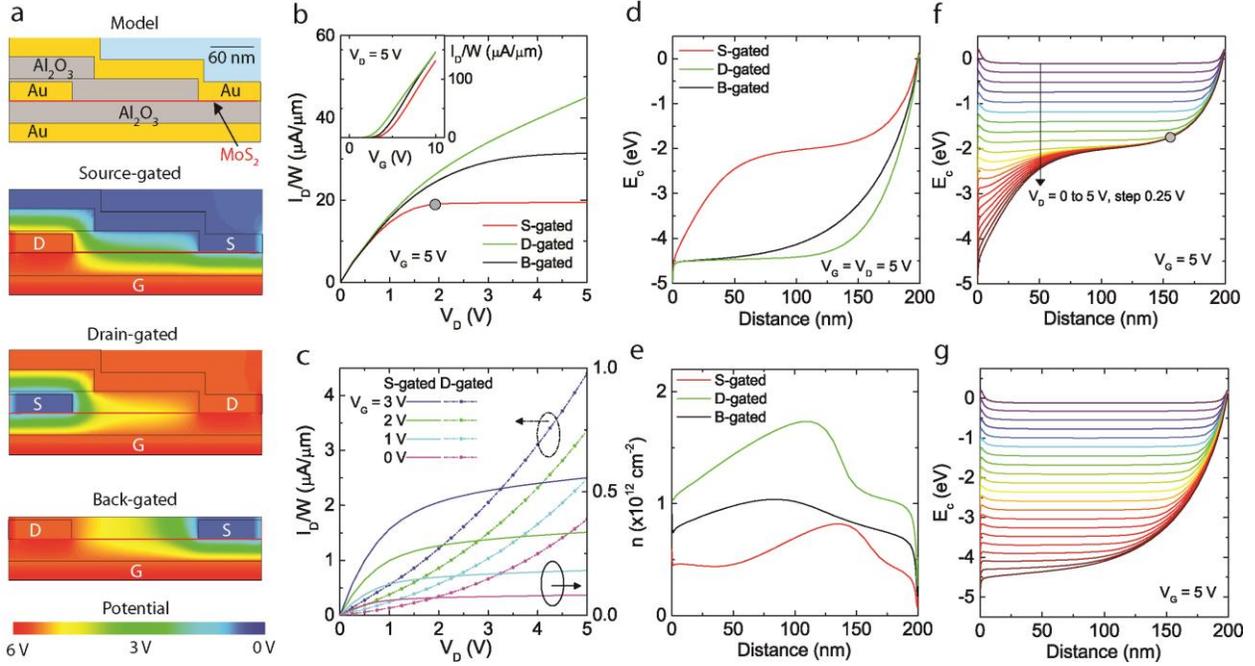

**Figure 2.** Electrostatics and charge transport simulations of SASC MoS$_2$ transistors. (a) Two-dimensional spatial map of the electrostatic potential of the SASC FET model ($L$ = 200 nm) in the source-gated (S-gated) and drain-gated (D-gated) biasing schemes with $V_D = V_G = 5$ V and $V_S = 0$ V. For comparison, the electrostatic potential map of a traditional back-gated (B-gated) FET ($L$ = 200 nm) is also shown where the left electrode is the drain. See Supporting Section S1.4 for modeling details. (b) Simulated output and transfer (inset) characteristics of the S-gated, D-gated, and B-gated FETs at $V_G = 5$ V and $V_D = 5$ V assuming Ohmic contacts. (c) Simulated output characteristics of the S-gated and D-gated devices with varying $V_G$, assuming more realistic thermionic emission and tunneling through the Schottky contacts that better reproduces the experimental data from Figure 1. (d) Calculated profile of the conduction band edge ($E_c$) as a function of distance along the channel for the S-gated, D-gated, and B-gated FETs from (a). (e) Profile of calculated carrier density ($n$) for the S-gated, D-gated, and B-gated FETs. $E_c$ and $n$ profiles for the D-gated case in (d) and (e) are inverted to keep the drain electrode on the left side

for facile comparison. (f,g) Evolution of the energy profiles as $V_D$ is increased from 0 to 5 V in steps of 0.25 V for the S-gated FET (f) and B-gated FET (g). Superior current saturation in the S-gated case is achieved by an additional pinch-off point near the source contact (< 50 nm from the electrode) at $V_D = 2$ V (gray dots in (b) and (f)).

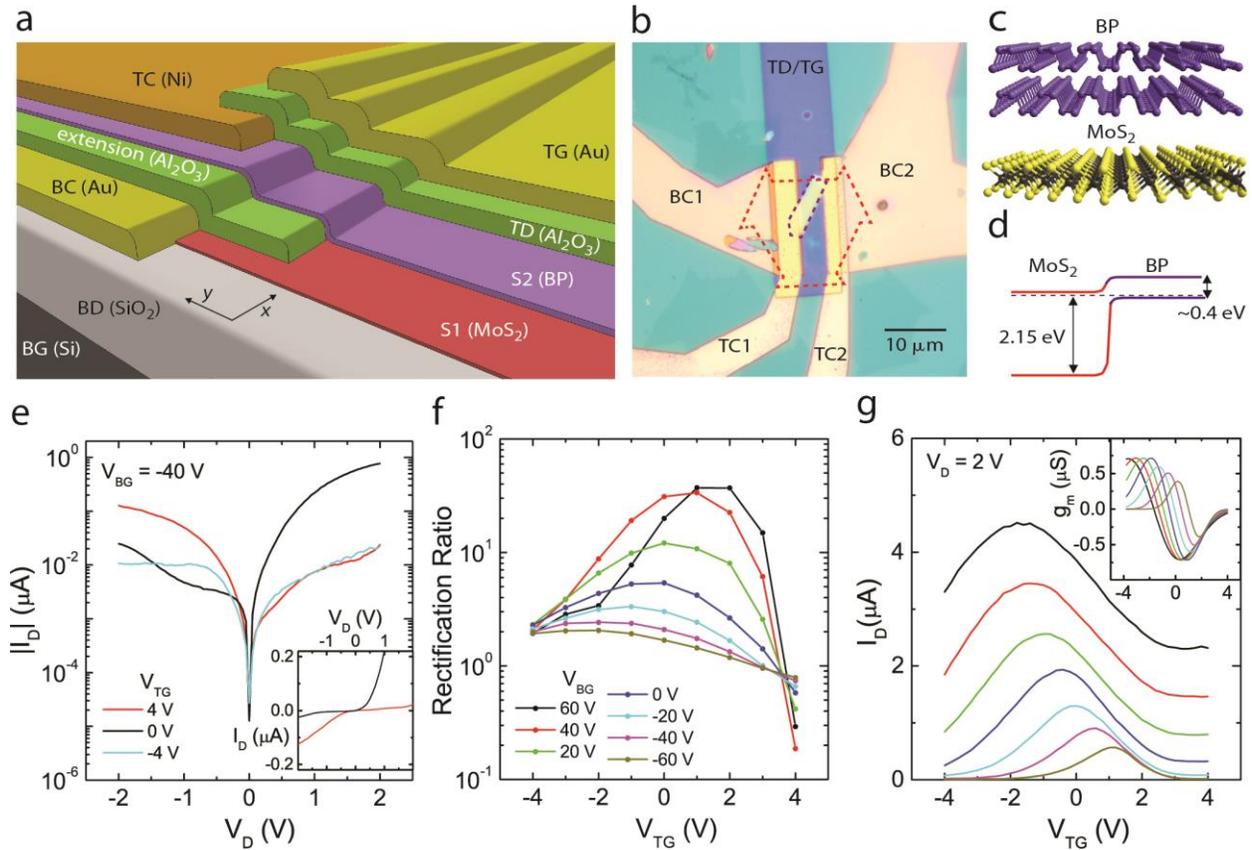

**Figure 3.** Self-aligned van der Waals heterojunction p-n diode. (a) Schematic of a self-aligned van der Waals heterojunction (vdWH) p-n diode using MoS$_2$ as semiconductor 1 (S1), few-layered black phosphorus (BP) as semiconductor 2 (S2), and ALD-grown 30 nm thick Al$_2$O$_3$ for both the dielectric extension and top gate dielectric (TD). Au and Ni serve as the bottom contact (BC) and top contact (TC) for minimum contact resistance with S1 and S2, respectively. A doped Si

substrate functions as global bottom gate (BG) with 300 nm thick thermal oxide as the bottom gate dielectric (BD). Offsets between layers in the y-direction are enabled by the self-aligned process. Artificial offsets in the x-direction are shown for clear visualization. (b) Optical micrograph of a BP-MoS$_2$ self-aligned p-n heterojunction taken before the last step of ALD growth/metallization for TD/TG through a patterned resist bilayer to assist visualization of different layers. The device consists of two p-n heterojunction diodes: one with BC1/TC1 electrode set on the left and the other with BC2/TC2 electrode set on the right. Outlines of the MoS$_2$ and BP flakes are shown by red and purple dashed lines, respectively. (c) Atomic structure of bilayer BP and monolayer MoS$_2$. (d) Energy band diagram of the BP-MoS$_2$ p-n heterojunction. (e) Current-voltage ($I_D$-$V_D$) characteristics of a BP-MoS$_2$ device at a bottom gate bias $V_{BG}$ = –40 V and top gate biases $V_{TG}$ = 4, 0, –4 V. Inset shows reversal of diode rectification at $V_{TG}$ = 0 (normal diode) and 4 V (Zener-like). The BP TC is the drain electrode (i.e., biased) and the MoS$_2$ BC is the source (i.e., grounded) throughout the vdWH measurements and simulations in Fig. 4. (f) Rectification ratio versus $V_{TG}$ for different values of V$_{BG}$ where the rectification ratio is defined as the ratio of the forward and reverse bias currents at $V_D$ = 2 V and –2 V, respectively. (g) $I_D$-$V_{TG}$ characteristics of the same device at different $V_{BG}$ values showing tunability of the anti-ambipolar response. The inset in (g) shows the variation in transconductance ($g_m$ = d$I_D$/d$V_{TG}$) obtained by fitting the $I_D$-$V_{TG}$ data with Gaussian profiles (see Supporting Fig. S2.13). The $V_{BG}$ legend in (f) also applies to the main plot and inset in (g).

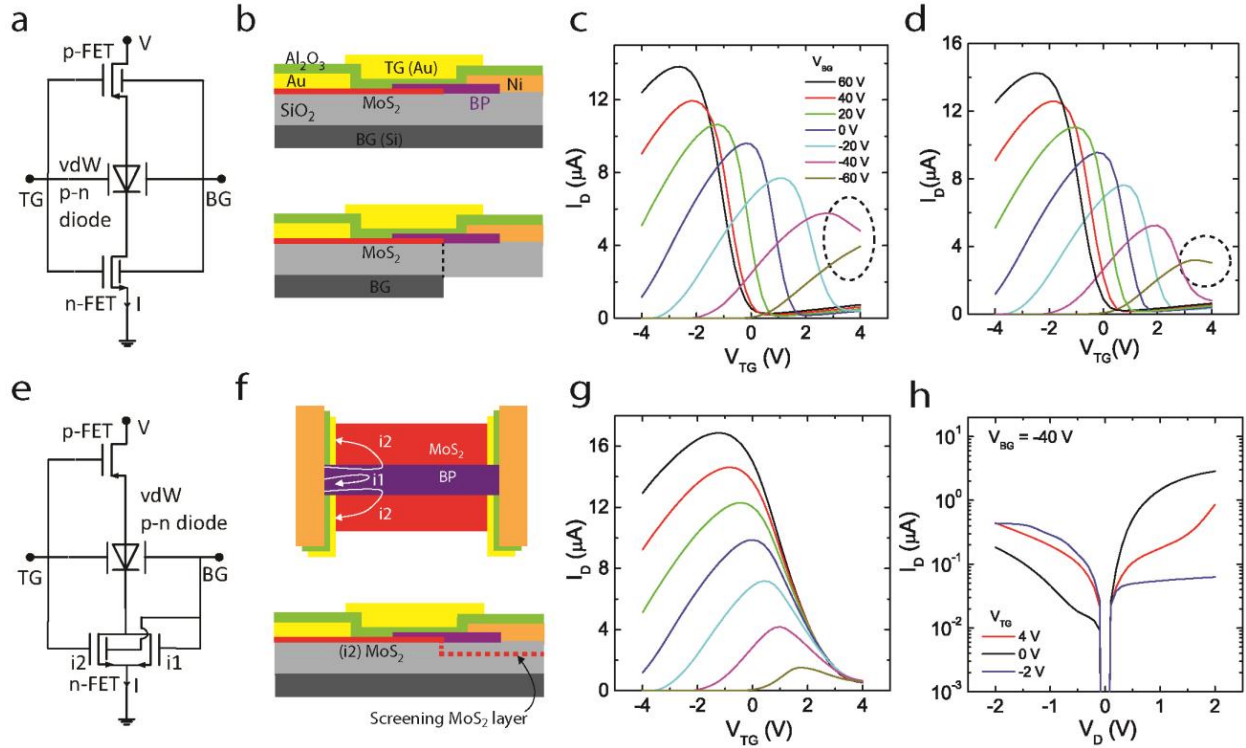

**Figure 4.** Device simulation for the dual-gated BP-MoS$_2$ p-n heterojunction. (a) Circuit diagram corresponding to the dual-gated vdWH where the p-n diode and both series transistors from the semiconductor extensions (p-type and n-type) are controlled by both of the gates. (b) Schematic of a dual-gated vdWH without (top) and with (bottom) alignment of the bottom gate with n-type MoS$_2$. See Supporting Fig. S2.10 for the exact model geometry. (c) Simulated $I_D$-$V_{TG}$ characteristics ($V_D$ = 2 V) of the vdWH at different $V_{BG}$ corresponding to the top panel in (b). The curves for $V_{BG}$ = –40 V and –60 V (dashed oval) show parasitic gating. (d) Simulated $I_D$-$V_{TG}$ characteristics ($V_D$ = 2 V) of the vdWH at different $V_{BG}$ corresponding to the bottom panel in (b). The curves for $V_{BG}$ = –60 V (dashed circle) show parasitic gating from the fringing field from the bottom gate despite the alignment with the MoS$_2$ edge. The $V_{BG}$ legend is shown in (c). (e) Circuit diagram corresponding to the dual-gated vdWH where the p-n diode and MoS$_2$ FET are controlled by both gates while the BP FET is controlled only by the top gate. (f) Top: Schematic of dual-

gated self-aligned vdWH (with TG/TD removed) showing two paths of current flow: i1 is from the BP to the region of MoS$_2$ directly underneath the BP; i2 is from the BP to the regions of MoS$_2$ that are not overlapping with the BP. i1 and i2 regions act as two transistors in parallel as shown in (e). Bottom: Equivalent planar model corresponding to the self-aligned vdWH where a screening MoS$_2$ layer (dashed red line) is embedded in the bottom dielectric, thereby isolating the BP transistor from the bottom gate. Transistor i1 in (e) and (f) is ignored for the planar model (see Supporting Fig. S2.10 for details). (g) Simulated $I_D$-$V_{TG}$ characteristics ($V_D = 2$ V) of the BP-MoS$_2$ vdWH at different $V_{TG}$ for the planar model in bottom panel in (f) showing no parasitic gating. $V_{BG}$ legend is shown in (c). (h) Simulated $I_D$-$V_D$ characteristics ($V_{BG} = -40$ V) of the BP-MoS$_2$ vdWH, which reveal a gradual reversal of rectification between $V_{TG} = 0$ V and $-2$ V, in qualitative agreement with the experimental data in Fig. 3e.